\documentclass[showpacs,floats,aps,pre,groupedaddress,showpacs,twocolumn,amsfonts,amssymb]{revtex4}
\usepackage{bm}
\usepackage{graphicx}

\newlength{\sepmod}
\begin{document}
\title{Random-cluster multi-histogram sampling for the $q$-state Potts model}

\author{Martin Weigel}
\email{weigel@itp.uni-leipzig.de}

\author{Wolfhard Janke}
\email{janke@itp.uni-leipzig.de}
\affiliation{Institut f\"ur Theoretische Physik,
  Universit\"at Leipzig, Augustusplatz 10/11, 04109 Leipzig, Germany}

\author{Chin-Kun Hu}
\email{huck@phys.sinica.edu.tw}
\affiliation{Institute of Physics, Academia Sinica, Nankang, Taipei, Taiwan 11529,
  Republic of China}

\date{July 06, 2001}

\begin{abstract}
  Using the random-cluster representation of the $q$-state Potts models we consider
  the pooling of data from cluster-update Monte Carlo simulations for different
  thermal couplings $K$ and number of states per spin $q$. Proper combination of
  histograms allows for the evaluation of thermal averages in a broad range of $K$
  and $q$ values, including non-integer values of $q$. Due to restrictions in the
  sampling process proper normalization of the combined histogram data is
  non-trivial. We discuss the different possibilities and analyze their respective
  ranges of applicability.
\end{abstract}


\maketitle

\section{Introduction}

During the last decade the question of how to make most efficient use of the data
sampled during a Monte Carlo (MC) simulation has received an increasing amount of
attention. The idea of {\em reweighting\/} \cite{ferrenberg:88a} of
time-series data from a single canonical simulation at a given fixed value of a
coupling parameter (i.e., most commonly temperature or magnetic field) to nearby
regions of the coupling-parameter space, allows for the analysis of thermal averages
as continuous functions of external parameters and thus a much more precise
determination of extremal, pseudo-critical points. As an extension of this, the
combination of data from simulations at {\em different\/} points in the
coupling-parameter space, commonly known as {\em multi-histogram technique\/}
\cite{ferrenberg:89a}, in principle allows to get accurate estimates for thermal
averages over a macroscopical region of couplings from a relatively small number of
simulation (that, however, generally has to be increased with the size of the
system).

The basic problem with collapsing data from different simulations is that of finding
the correct relative normalization of the single histograms. Consider the sampled
energy histogram $\hat{H}_{K_i}(E)$ of, e.g., an Ising model simulation at the
coupling $K_i = J\beta_i$, consisting of $N$ energy-measurements. The thermal average
of an observable $A(E)$ at $K_i$ is just given by the time-series average in the
importance-sampling scheme and thus insensitive to the value of the partition
function at that point. Combining two histograms, however, amounts to the combination
of the temperature-independent expressions
\begin{equation}
  Z_{K_i} \hat{H}_{K_i}(E)/N\,e^{K_i E}
  \label{simple_density}
\end{equation}
for different simulations $i$, where the partition function $Z_{K_i}$ appears as a
normalization constant. Thus, for correct relative normalization of the histograms to
be combined, one has to know the ratio of partition functions
$Z_{K_{i_1}}/Z_{K_{i_2}}$ or, equivalently, differences in the free energy densities
$f_{K_{i_1}}-f_{K_{i_2}}$ at the simulated couplings $K_i$. In
Ref.~\cite{ferrenberg:89a} this problem has been solved by an iterative solution of
self-consistency equations for the free energy differences at adjacent simulation
couplings $K_i$.

Since the combination given in Eq.~(\ref{simple_density}) is nothing but an estimator
for the density of (energy) states $\Omega(E)$, multi-histograming data analysis
amounts to estimating the density of states of the variable that is thermodynamically
conjugate to the considered coupling parameter. Going to the random-cluster
representation of the Potts model, i.e., its interpretation as correlated percolation
model \cite{fortuin:72a,coniglio:80a,hu:84a}, the relevant density of states is given
by the number $g(b,n)$ of bond configurations with $b$ bonds and $n$ clusters on the
lattice. Apart from gaining control over {\em two\/} parameters, the thermal coupling
$K$ and the number of states $q$, this language suggests the use of cluster
estimators for thermal averages like correlation functions, which are known to yield
a variance-reduction in certain situations \cite{wolff:88a}. One of us
\cite{hu:92b,hu:94a} has proposed a multi-histogram technique for the $q$-state Potts
model and simulations at different temperatures making use of the sampling of cluster
decompositions of the lattice as they occur in the Swendsen-Wang cluster-update
algorithm \cite{swendsen-wang:87a}. There, the relative normalization of the
individual histograms at couplings $K_i$ is accomplished by making use of the known
absolute number of configurations with $b$ active bonds on the lattice, which is just
given by the binomial ${{\cal E} \choose b}$, ${\cal E}$ being the total number of
bonds of the lattice. While this method appears advantageous at first sight and gives
nice results from the cases of percolation ($q\rightarrow 1$) \cite{hu:92a} and the
Ising model ($q=2$) \cite{hu:94a}, we find that this procedure is not the best choice
of normalization for simulations of Potts models with $q$ larger than 2 or 3 and
propose a different approach for normalization to circumvent this problem.

The outline of the paper is as follows. In Section II we restate the multi-histogram
approach of Ref.~\cite{hu:94a} in the random-cluster representation (``RC
histograming''), which was originally formulated for simulations at fixed $q$ only,
and generalize it to simulations of multiple $q$ values. Applying it to the $q=10$
Potts model in two dimensions we find large deviations from the expected results. As
an alternative, in Section III we propose an adaptively normalized RC
multi-histograming ansatz. We discuss details of its implementation and present a
comparative reweighting analysis for the $q=10$ case.  For energy-related observables
we also compare histograming in the random-cluster language to the sampling of the
{\em energy\/} density of states in the well-known framework of histograming in the
energy/magnetization language (``EM histograming'') \cite{ferrenberg:89a}.  Comparing
both methods, in Section IV we track down the observed deviations with the first
ansatz to be a result of the application of the above mentioned normalization
condition. This problem can thus be resolved by the second ansatz.
Finally, Section V contains our conclusions.

\section{RC Histograms and absolute normalization}

Consider the Hamiltonian of the $q$-state Potts model in zero magnetic field,
\begin{equation}
  {\cal H} = -J \sum_{\langle i,j\rangle}\delta(\sigma_i,\sigma_j),
  \;\;\;\;\sigma_i = 1,\ldots,q,
  \label{potts_hamilton}
\end{equation}
on a general graph ${\cal G}$ with ${\cal N}$ sites and ${\cal E}$ bonds.
Transforming to the random-cluster representation \cite{fortuin:72a}, the partition
function becomes
\begin{equation}
  Z \equiv \sum_{\{\sigma_i\}} e^{K\sum_{\langle i,j\rangle}\delta(\sigma_i,\sigma_j)}
  = \sum_{{\cal G}' \subseteq {\cal G}} (e^K-1)^{b({\cal G}')}\,q^{n({\cal G}')},
  \label{random_cluster}
\end{equation}
where the sum runs over all bond configurations ${\cal G}'$ on the graph (subgraphs),
and $K=\beta J$ denotes the thermal coupling. Notice that the formulation
(\ref{random_cluster}) in contrast to that of Eq.~(\ref{potts_hamilton}) allows
for a natural continuation of the model to {\em non-integer\/} values of the parameter $q$.
Using the subgraph expansion of the $q$-state Potts in external field, one of us
\cite{hu:84a} has shown that the $q$-state Potts model can be considered as a
bond-correlated percolation model (BCPM) with bond occupation probability
$p=1-e^{-K}$. Eq.~(\ref{random_cluster}) can be rewritten as:
\begin{eqnarray}
  Z_{p,q}({\cal G}) & = & e^{K{\cal E}} \sum_{{\cal G}' \subseteq {\cal G}}
  p^{b({\cal G}')}(1-p)^{{\cal E}-b({\cal G}')}\,q^{n({\cal G}')} \nonumber \\
  & = & e^{K{\cal E}} \sum_{b=0}^{{\cal E}}
  \sum_{n=1}^{{\cal N}}g(b,n)\,p^b\,(1-p)^{{\cal E}-b}\,q^n,
  \label{partition}
\end{eqnarray}
where $g(b,n)$ denotes the number of subgraphs of ${\cal G}$ with $b$ activated
bonds and $n$ clusters resulting therefrom. This purely combinatorial quantity
corresponds to the density of states of the BCPM.

The Swendsen-Wang cluster-update algorithm generates bond configurations drawn from
the equilibrium canonical distribution of this model. Thus, the probability for the
occurrence of a subgraph with $b$ bonds and $n$ clusters is given by
\begin{equation}
  P_{p,q}(b,n) = W_{p,q}^{-1}({\cal G})\,g(b,n)\,p^b\,(1-p)^{{\cal E}-b}\,q^n,
\end{equation}
which in turn is the expectation value of the normalized sampled histogram of bond
configurations, i.e., $P_{p,q}(b,n) = \langle\hat{H}_{p,q}(b,n)/N\rangle$, where $N$
denotes the length of the time series of measurements. Here, we separated the common
factor $\exp(K{\cal E})$ from the partition function:
\begin{equation}
  Z_{p,q}({\cal G}) = e^{K{\cal E}}\,W_{p,q}({\cal G}).
\end{equation}
An estimator for the density of states $g(b,n)$ is therefore given by
\begin{equation}
  \hat{g}(b,n) = W_{p,q}({\cal G}) \frac{\hat{H}_{p,q}(b,n)}{p^b\,(1-p)^{{\cal
        E}-b}\,q^n\,N}.
  \label{first_nt_est}
\end{equation}
Since the reduced partition function $W_{p,q}({\cal G})$ is {\em a priori\/} unknown,
the correct normalization of this estimator is not known at the beginning.  Probably
the most obvious way of fixing the normalization would be to estimate the reduced
partition function $W_{p,q}({\cal G})$ directly from Eq.~(\ref{first_nt_est}). One
can do better than that, however, by considering the accumulated density $g(b)$, which
is obviously just a binomial \cite{hu:92b},
\begin{equation}
  g(b) = \sum_n g(b,n) = {{\cal E} \choose b}.
  \label{sum_rule}
\end{equation}
Imposing this restriction on the estimate $\hat{g}(b,n)$ also, one arrives at
\begin{equation}
  \hat{C}_{p,q}(b) \equiv \frac{\hat{W}_{p,q}({\cal G})}{p^b\,(1-p)^{{\cal E}-b}N} = 
    \frac{{{\cal E} \choose b}}{\sum_n\hat{H}_{p,q}(b,n)\,q^{-n}},
    \label{sum_rule_data}
\end{equation}
so that the absolute values of $\hat{g}(b,n)$ are now fixed by ${\cal E}$
independent normalization conditions, one for each number of active bonds $b$. Thus
we have the following estimate for the density of states \cite{hu:94a}:
\begin{equation}
  \hat{g}(b,n) = \hat{C}_{p,q}(b)\,\hat{H}_{p,q}(b,n)\,q^{-n}.
  \label{single_nt_est}
\end{equation}

Now, we want to combine the estimates $\hat{g}^{(i)}(b,n)$ from several simulations
at different parameters $(p_i,q_i)$, i.e., we want to do multi-histograming in both
parameters, $p$ and $q$. Then, we have
\begin{equation}
  \hat{g}(b,n) = \sum_i \alpha_i(b,n)\,\hat{g}^{(i)}(b,n),\hspace{3mm}\sum_i
  \alpha_i(b,n) = 1.
\end{equation}
Since we want to minimize the variance $\hat{\sigma}^2[\hat{g}]$ of the final
estimate and the different simulations are statistically independent, the correct
choice of the weights $\alpha_i$ obviously is given by
\begin{equation}
  \alpha_i(b,n) = \frac{1/\hat{\sigma}^2[\hat{g}^{(i)}(b,n)]}{\sum_i
    1/\hat{\sigma}^2[\hat{g}^{(i)}(b,n)]}.
\end{equation}
From Eq.~(\ref{single_nt_est}) we have
\begin{eqnarray}
  \hat{\sigma}^2[\hat{g}^{(i)}(b,n)] & = & \hat{C}^2_{p_i,q_i}(b)\,q_i^{-2n}\,
  \hat{\sigma}^2[\hat{H}_{p_i,q_i}(b,n)] \nonumber \\
  & \approx & \hat{C}^2_{p_i,q_i}(b)\,q_i^{-2n}\,\hat{H}_{p_i,q_i}(b,n),
  \label{sigma_nt}
\end{eqnarray}
such that the variance-optimized estimate for $g(b,n)$ becomes
\begin{equation}
  \hat{g}(b,n) = \frac{{{\cal E} \choose b}\sum_i\sum_{\mu}\hat{H}_{p_i,q_i}(b,\mu)\,q_i^{-\mu}\,q_i^n}
  {\sum_j{[\sum_{\nu}\hat{H}_{p_j,q_j}(b,\nu)\,q_j^{-\nu}]}^2\,q_j^{2n}\,
    [\hat{H}_{p_j,q_j}(b,n)]^{-1}}.
  \label{nt_abs}
\end{equation}
In writing this expression we allow for several approximations: first, we treat
$\hat{C}_{p_i,q_i}(b)$ as a parameter in Eq.~(\ref{sigma_nt}) instead of taking its
own variance into account; this is justified by the clear suppression of variance of
this quantity as a sum as compared to the the variance of its summands
$\hat{H}_{p_i,q_i}(b,n)$. Secondly, we take
$\hat{\sigma}^2[\hat{H}(b,n)]=\hat{H}(b,n)$, i.e., we treat the individual bins
$(b,n)$ as independently distributed according to an uncorrelated $1/N$ statistics,
which will in general not be exactly fulfilled. Since those assumptions only affect
the variance of the final estimate, however, and do not introduce a bias, we consider
them justified. Finally, we do not take autocorrelations between successive
measurements $(b,n)$ into account, i.e., we assume here and in the following that
measurements in the sampling process are taken with a frequency around
$1/\tau_\mathrm{int}$, where $\tau_\mathrm{int}$ denotes the integrated
autocorrelation time, resulting in an effectively uncorrelated time series.

\begin{figure}[!tb]
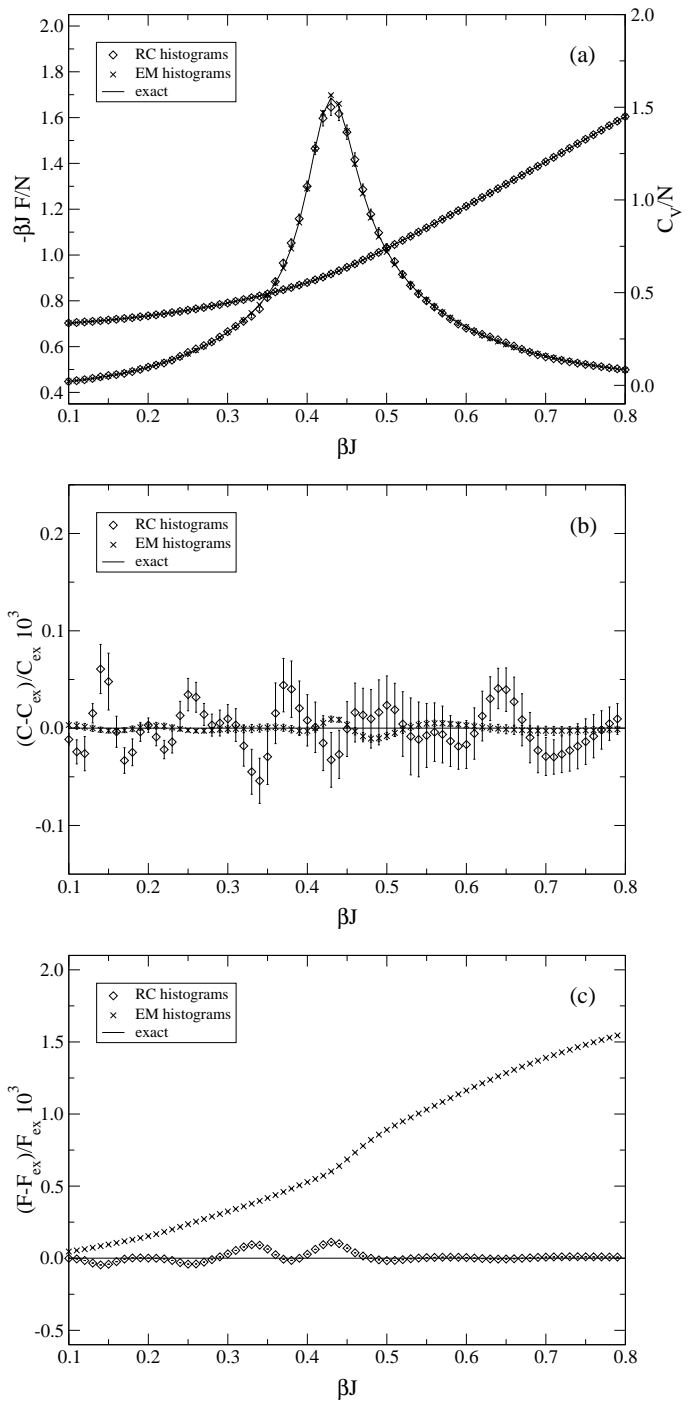

  \unitlength8.75cm
  \begin{picture}(1,0.7)
    \put(-0.00886287625418057, -0.0313628762541806){\includegraphics[angle=0,clip=true,
      width=8.91753762541806cm,keepaspectratio=true]{fig1a.eps}}
  \end{picture}

  \vspace*{1mm}

  \unitlength8.75cm
  \begin{picture}(1,0.7)
    \put(-0.0173913043478261, -0.0313628762541806){\includegraphics[angle=0,clip=true,
      width=8.90510033444816cm,keepaspectratio=true]{fig1b.eps}}
  \end{picture}

  \vspace*{1mm}

  \unitlength8.75cm
  \begin{picture}(1,0.7)
    \put(-0.0173913043478261, -0.0313628762541806){\includegraphics[angle=0,clip=true,
      width=8.90510033444816cm,keepaspectratio=true]{fig1c.eps}}
  \end{picture}

  \caption{Results from the EM and RC multi-histogram analyses of
    time series from $9$ cluster update simulations of the two-dimensional $q=2$
    Potts model on a ${\cal N}=16^2$ lattice. (a) Free energy density (left scale)
    and specific heat (right scale) as a function of the coupling $K=\beta J$ as
    compared to the exact solution of Ref.~\cite{ferdinand:69a}. (b) Relative
    deviation of the results for the specific heat from the exact solution for both
    methods. (c) Relative deviation for the free energy. All results shown are
    re-scaled from the $q=2$ Potts model to the Ising model formulation to fit the
    results from Ref.~\cite{ferdinand:69a}.}
  \label{ising_hu_sf}
\end{figure}

\begin{figure}[!tb]
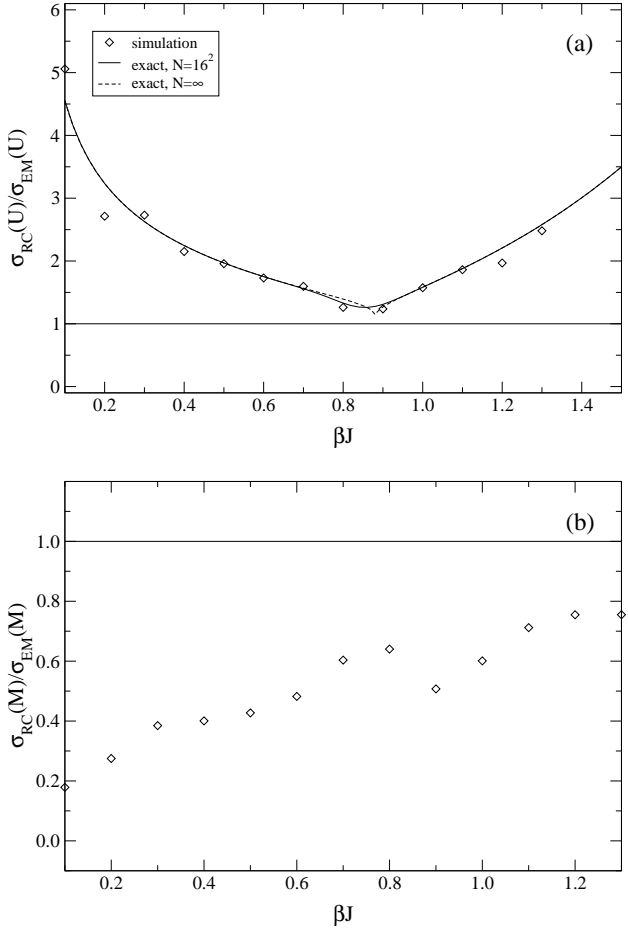

  \unitlength8.75cm
  \begin{picture}(1,0.7)
    \put(-0.0117056856187291, -0.0313628762541806){\includegraphics[angle=0,clip=true,
      width=8.19617474916388cm,keepaspectratio=true]{fig2a.eps}}
  \end{picture}

  \vspace*{2mm}

  \unitlength8.75cm
  \begin{picture}(1,0.7)
    \put(-0.0117056856187291, -0.0313628762541806){\includegraphics[angle=0,clip=true,
      width=8.24592391304348cm,keepaspectratio=true]{fig2b.eps}}
  \end{picture}
  \caption{Ratio of standard deviations for estimates of (a) the internal energy $U$
    and (b) the spontaneous magnetization $\tilde{M}$ of the $q=2$ Potts model on a
    ${\cal N}=16^2$ lattice and RC and EM histogram analyses as a function of the
    coupling $K=\beta J$. The variances are estimated by a ``jackknife'' time series
    analysis \cite{efron:82}. The solid line of (a) shows the exact result of
    Eq.~(\ref{energy_error_eq}) and Ref.~\cite{ferdinand:69a}, the dashed line that
    for ${\cal N}\rightarrow\infty$ from Ref.~\cite{onsager:44}. In (b) we use the
    definition (\ref{m_unbroken}) for $K<0.8$ and (\ref{m_broken}) for $K\ge 0.8$.}
  \label{energy_error}
\end{figure}

\begin{figure}[!tb]
  \unitlength8.75cm
  \begin{picture}(1,0.7)
    \put(-0.00459866220735784, -0.0313628762541806){\includegraphics[angle=0,clip=true,
      width=8.28323578595318cm,keepaspectratio=true]{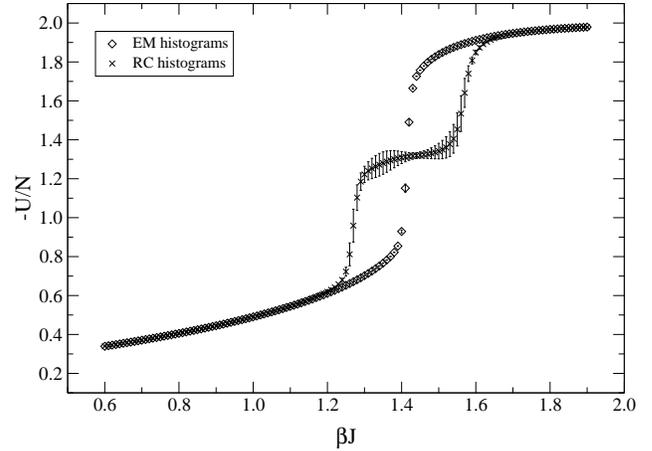}}
  \end{picture}
  \caption{Internal energy of the two-dimensional $q=10$ Potts model on a ${\cal N}=16^2$ square lattice
    with periodic boundary conditions as given by the random-cluster (RC) and energy
    representation (EM) multi-histogram analyses from simulations for different
    thermal couplings $K$. The transition point of the infinite system is given by
    $K_t = \ln(1+\sqrt{10})\approx 1.426$ \cite{wu:82a}.}
  \label{q10_hu_sf}
\end{figure}

From the partition function Eq.~(\ref{partition}) we infer the following
cluster-language estimators for the free energy density $f=F/{\cal N}$, the per-site
internal energy $u=U/{\cal N}$ and specific heat $c_v=C_v/{\cal N}$,
\begin{eqnarray}
  \hat{f} & = & -\frac{1}{K{\cal N}}\,\ln\hat{Z}_{p,q}({\cal G}), \nonumber \\
  \hat{u} & = & -\frac{1}{p{\cal N}}\,{\langle b\rangle}_{\hat{g}}, \nonumber \\
  \hat{c}_v & = & \frac{K^2}{p^2{\cal N}}\,\left[{\langle(b-\langle
      b\rangle_{\hat{g}})^2\rangle}_{\hat{g}}
    -(1-p){\langle b\rangle}_{\hat{g}}\right],
  \label{hu_thermal}
\end{eqnarray}
cf.\ Appendix A. Here, the estimated expectation value of an observable $O(b,n)$ is
defined as
\begin{eqnarray}
  {\langle O(b,n)\rangle}_{\hat{g}} & \equiv & \hat{Z}_{p,q}^{-1}({\cal
  G})\,e^{K{\cal E}}\,\sum_{b=0}^{\cal E}
  \sum_{n=1}^{\cal N} \hat{g}(b,n) \nonumber \\
  & & \times\;p^b\,(1-p)^{{\cal E}-b}\,q^n\,O(b,n).
\end{eqnarray}

For the evaluation of magnetic observables one has to distinguish percolating
clusters, denoted by indices $\pi$, from non-percolating, finite clusters,
denoted by indices $\phi$. Let $\{c(G')\}$ be the set of clusters of a subgraph $G'$
of the lattice and $n_c(G')$ the number of sites in cluster $c$ of $G'$. Then,
consider the following microcanonical averages:
\begin{equation}
  m_1^\pi(b,n) \equiv \frac{1}{{\cal N}g(b,n)}\sum_{\stackrel{{\cal G}' \subseteq {\cal G}, b(G')=b,}{
    n(G')=n}}\sum_{\{c^\pi(G')\}} n_c^\pi(G'),
\end{equation}
i.e., the average number of sites in percolating clusters for subgraphs with $b$
active bonds and $n$ clusters,
\begin{equation}
  m_2^\pi(b,n) \equiv \frac{1}{{\cal N}^2g(b,n)}\sum_{\stackrel{{\cal G}' \subseteq {\cal
  G}, b(G')=b,}{n(G')=n}} \left[\sum_{\{c^\pi(G')\}} n_c^\pi(G')\right]^2,
\end{equation}
i.e., the mean square number of sites in non-percolating clusters for those
subgraphs, and
\begin{equation}
  m_3^\phi(b,n) \equiv \frac{1}{{\cal N}g(b,n)}\sum_{\stackrel{{\cal G}' \subseteq {\cal
  G}, b(G')=b,}{n(G')=n}} \sum_{\{c^\phi(G')\}} [n_c^\phi(G')]^2,
\end{equation}
i.e., the mean squared sum of the size of non-percolating clusters. These
microcanonical averages obviously can be estimated by adding $\sum_{\{c^\pi(G')\}}
n_c^\pi(G')$ to $\hat{M}_{1,p,q}^\pi(b,n)$ for each bond configuration $G'$ with
$\{b(G')=b,n(G')=n\}$, where $n_c^\pi(G')$ should be taken $0$ for non-percolating
bond configurations, adding $[\sum_{\{c^\pi(G')\}} n_c^\pi(G')]^2$ to
$\hat{M}_{2,p,q}^\pi(b,n)$, and adding $\sum_{\{c^\phi(G')\}} [n_c^\phi(G')]^2$ to
$\hat{M}_{3,p,q}^\phi(b,n)$ for each such observed configuration. Then, if we define
\begin{equation}
  \begin{array}{rcl}
    \hat{H}(b,n) & = & \sum_{p_i,q_i}\hat{H}_{p_i,q_i}(b,n), \\
    \hat{M}_1^\pi(b,n) & = & \sum_{p_i,q_i} \hat{M}_{1,p_i,q_i}^\pi(b,n), \\
    \hat{M}_2^\pi(b,n) & = & \sum_{p_i,q_i} \hat{M}_{2,p_i,q_i}^\pi(b,n), \\
    \hat{M}_3^\phi(b,n) & = & \sum_{p_i,q_i} \hat{M}_{3,p_i,q_i}^\phi(b,n), \\
  \end{array}
  \label{histo_sum}
\end{equation}
we have the following estimates for $m_1^\pi(b,n)$, $m_2^\pi(b,n)$, and $m_3^\phi(b,n)$,
\begin{equation}
  \hat{m}_{1/2/3}^{\pi/\phi}(b,n) = \frac{\hat{M}_{1/2/3}^{\pi/\phi}(b,n)}{{\cal N}\hat{H}(b,n)},
  \label{mest}
\end{equation}
which, finally, result in the following expressions for the (zero-field)
magnetization $\tilde{m}$ and the magnetic susceptibility $\tilde{\chi}$,
\begin{eqnarray}
  \hat{\tilde{m}} & = & \frac{q-1}{q} \left\langle\hat{m}_1^\pi(b,n)\right\rangle_{\hat{g}}+\frac{1}{q},
    \nonumber \\
    \hat{\tilde{\chi}} & = & {\cal
    N}\left(\frac{q-1}{q}\right)^2\left[\left\langle m_2^{\pi}
      \right\rangle_{\hat{g}}-\left\langle m_1^{\pi}
    \right\rangle_{\hat{g}}^2\right] \nonumber \\
    & & +\frac{q-1}{q^2}\left\langle m_3^{\phi}\right\rangle_{\hat{g}},
\end{eqnarray}
cf.\ Appendix A. Note, that we simply add up histograms from different simulations in
Eqs.  (\ref{histo_sum}) and (\ref{mest}) without using any reweighting factors in $p$
and $q$. This is correct since the conditional probability of the occurrence of, say,
a given number of sites in percolating clusters in a subgraph with $b$ active bonds
and $n$ clusters does no longer depend on $p$ and $q$. The order parameter
$m$ of the Potts model is given by \cite{wu:82a}
\begin{equation}
  m = \frac{q\tilde{m}-1}{q-1},
\end{equation}
and the corresponding, rescaled susceptibility is $\chi=[(q-1)/q]^2\tilde{\chi}$.

As a first comparative test for the method we performed a Swendsen-Wang cluster MC
simulation for the Ising model case ($q=2$) on a small, simple-cubic ${\cal N}=16^2$
lattice with periodic boundary conditions. We gathered histograms from 9 different
simulations at the couplings $K_i = 0.1, 0.2, \ldots, 0.8$ and $K = K_c =
\frac{1}{2}\ln(1+\sqrt{2})$, where the couplings are given in the language of the
Ising model in this case, i.e., are half of the couplings of the corresponding $q=2$
Potts model.  Each run sampled $2^{17} = 131\,072$ bond configurations resulting in
corresponding time series of $(b,n)$ and of the energy/magnetization pairs $(E,M)$
for comparison with the EM histograming method. Thus, any differences in the results
must be solely due to the method of data analysis, the underlying simulation data
being exactly identical. For the EM histograms throughout this paper we use a
multi-histogram analysis according to Ref.~\cite{ferrenberg:89a} very similar to that
presented for the RC histograms in Sec. III. Figure~\ref{ising_hu_sf} shows the
results in comparison to the exact expression for $F$ and $C_v$ on square lattices as
given by Kaufman \cite{kaufman:49a} and analysed by Ferdinand and Fisher
\cite{ferdinand:69a}.  Statistical errors for both analysis schemes were evaluated
using the ``jackknife'' error estimation technique \cite{efron:82}. The relative
deviations $(\hat{C}_v-C_v)/C_v$ from the exact result are noticeably larger for the
RC histogram analysis, however in agreement with statistical errors in both cases,
cp.\ Fig.~\ref{ising_hu_sf}(b). The same holds true for the internal energy $U$ given
by the different estimates, which is not shown in Fig.~\ref{ising_hu_sf}. Note that
for the energy related observables from Eq.~(\ref{hu_thermal}) only the limiting
distribution $P_{p,q}(b)$ is needed, which, in general, has a different width than
the distribution of energies $P_{p,q}(E)$, thus leading to different variances.

In fact, by inspection of the random-cluster expression for the specific heat
Eq.~(\ref{hu_thermal}) and comparison with its definition in the energy language as
$C_v = K^2(\langle E^2\rangle-\langle E\rangle^2)$ we can infer the following
relation between the variances of energy estimates in the RC and EM schemes:
\begin{equation}
  \frac{\sigma^2_{RC}(U)}{\sigma^2_{EM}(U)} = 1-K^2\,\frac{1-p}{p}\,
  \frac{U}{C_v}\,\ge\,1.
  \label{energy_error_eq}
\end{equation}
Thus, energy estimates from RC histograms are always less precise than those from EM
histograms, regardless of the temperature. Figure \ref{energy_error}(a) shows the
ratio of jackknife-estimated variances of the two different estimates of internal
energy, compared to the result from Eq.~(\ref{energy_error_eq}) with the exact
expressions for $U$ and $C_v$ for the $q=2$ case inserted \cite{ferdinand:69a}. Note,
that from Fig.~\ref{energy_error}(a) this quantity seems to have extremely small
finite-size corrections. As a reminder, this shows clearly that cluster estimators
are not always improved estimators \cite{wolff:90a}, but sometimes ``deteriorated
estimators''.  Note, however, that this effect will decrease with increasing number
of states $q$, at least in the transition region, since the singularity in $C_v$
sharpens in this limit, whereas the energies $U$ always stay in the range $0\le
U/{\cal N}\le 2$. For the $q=10$ model is has been observed that at the transition
point $P_{p,q}(b)$ is almost indistinguishable from $P_{p,q}(E)$, when suitably
rescaled \cite{wj:95a}. The minimum of the exact curve of Fig.~\ref{energy_error} at
the critical point is somewhat in contrast to the usual notion that cluster
estimators work best off criticality \cite{wolff:88a}; this result, however, applies
to the spin-spin correlation function at medium and long distances and to magnetic
observables like the susceptibility, which is the integral of the correlation
function, whereas the internal energy $U$ constitutes the extreme short distance
limit of this quantity. For the magnetic observables $m$ and $\chi$ the situation is
reversed, the variance of the RC estimators being strongly reduced as compared to the
EM estimators, cp.\ Fig.~\ref{energy_error}(b). Note that in contrast to the EM case,
the RC estimators provide a single consistent definition of $m$ and $\chi$ for both,
the broken and unbroken phases, cf.\ Appendix A.

For the free energy, on the other hand, deviations for the RC method are by far
smaller than those of the EM method, cp.\ Fig.~\ref{ising_hu_sf}(c). Moreover,
deviations are not covered by statistical errors in the latter case, a fact we will
comment on later in the next Section. In the EM case $F$ is being fixed by making
contact with the non-interacting limit $K = 0$ resp. $p = 0$, where
\begin{equation}
  Z_{p=0,q}({\cal G}) = \sum_E \Omega(E) = q^{\cal N},
  \label{omega_norm}
\end{equation}
so that $-KF(p=0)/{\cal N} = \ln q$. This equation corresponds to the normalization
condition Eq.~(\ref{sum_rule}). It is obvious that having a normalization condition
for each number $b$ of active bonds and therefore, implicitly, for each
(microcanonical) temperature in the RC case allows for accurate estimation of the
free energy even far away from $K=0$, whereas for EM histograms the results
deteriorate with the distance from the only normalization point $K = 0$. Thus, for
sampling free energies the RC multi-histogram technique normalized by
Eq.~(\ref{sum_rule}) seems to be a good choice.

As a slightly less trivial example we performed simulations for the $q=10$ Potts
model on the same lattice, which exhibits a strongly first-order phase transition. It
is well known that cluster algorithms are not efficient to reduce the
``super-critical'' (exponentially strong) slowing down of the local MC dynamics at
first-order transitions. For the small lattice under consideration, however,
autocorrelation times are still quite moderate, so that one gets reliable results
without having to resort to more sophisticated methods like multi-canonical
simulations \cite{berg:92b,wj:95a}. We gathered data from 11 single-histogram
simulations at couplings $K_i = 0.8, 0.9, \ldots, 1.8$ with $2^{20}=1\,048\,576$
measurements each. Figure \ref{q10_hu_sf} shows the quite astonishing results for the
internal energy from this simulation data using the analyses in the RC and EM
languages, respectively. Naturally, we do not have exact results to compare with in
this case; nevertheless, the results from the EM analysis are completely in agreement
with our expectations and also well compatible with results from previous simulations
\cite{wj:97a}. So, obviously, the results from the RC histogram analysis are
strikingly wrong --- and in a way that is clearly not covered by the present
statistical errors. Obviously, the results for the specific heat, which are not
shown, look even worse, with a pronounced, unphysical double-peak resulting from the
deviations in internal energy shown in Fig.~\ref{q10_hu_sf}.

\begin{figure}[!tb]
  \unitlength8.75cm
  \begin{picture}(1,0.7)
    \put(-0.00459866220735784, -0.0313628762541806){\includegraphics[angle=0,clip=true,
      width=8.13398829431438cm,keepaspectratio=true]{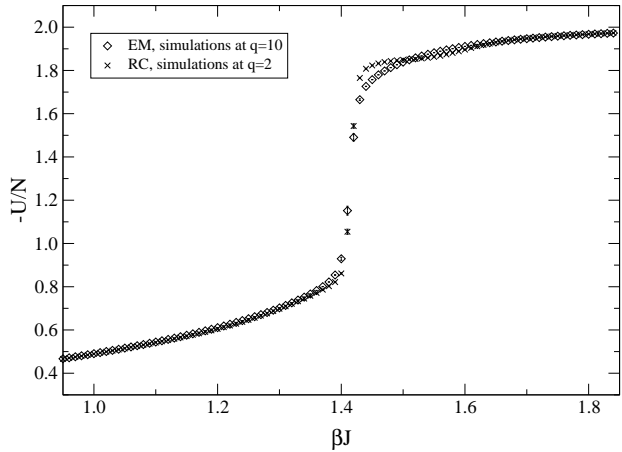}}
  \end{picture}
  \caption{Internal energy of the $q=10$ Potts model on a ${\cal N}=16^2$ square lattice,
    reweighted from the $q=2$ model simulations shown in Fig.~\ref{ising_hu_sf} using
    multiple RC histograms according to Eq.~(\ref{nt_abs}). The $q=10$ results from
    the EM multi-histogram analysis are shown for comparison.}
  \label{q10_ising_hu_sf}
\end{figure}

Since as one of its major strengths in the RC approach we have the possibility of
reweighting in the parameter $q$ also, as a first clue to the reason for this
conspicuous failure we show the outcome of using the $q=2$ simulation data from above
for determining the internal energy of the $q=10$ case, cp.\ 
Fig.~\ref{q10_ising_hu_sf}. The agreement with the direct EM analysis of the $q=10$
simulations is remarkably good considering the large distance in $q$ between the
simulation and analysis points. Comparing Figs.~\ref{q10_hu_sf} and
\ref{q10_ising_hu_sf} it is quite natural to suspect that the application of the
normalization condition (\ref{sum_rule}) is not a proper choice for simulation data
from larger $q$ models.

\begin{figure}[!t]
  \unitlength8.75cm
  \begin{picture}(1,0.7)
    \put(-0.00744147157190635, -0.0270986622073578){\includegraphics[angle=0,clip=true,
      width=8.15886287625418cm,keepaspectratio=true]{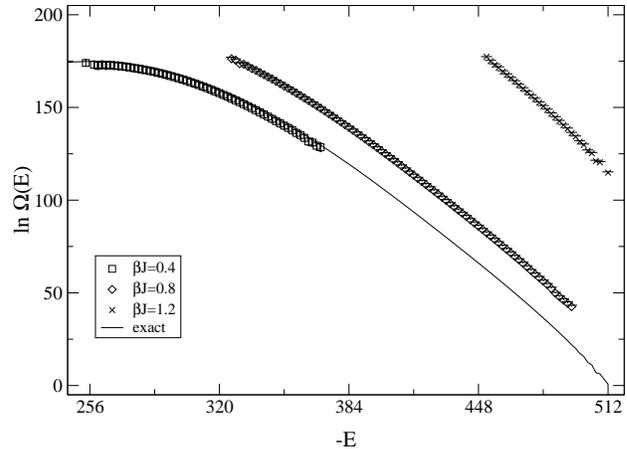}}
  \end{picture}
  \caption{Density of states for the $q=2$ Potts model in two dimensions on a
    ${\cal N}=16^2$ lattice from single-histogram cluster-update simulations at
    coupling $K=0.4$, 0.8 and 1.2 using the estimator Eq.~(\ref{omega_est}). The
    solid line shows the exact result of Ref.~\cite{beale:96a}.}
  \label{omega_fig}
\end{figure}

\section{RC Histograms and adaptive normalization}

To understand this normalization problem let us shortly go back to the sampling of
the energy density of states $\Omega(E)$ for the case of the two-dimensional $q=2$
(Ising) model.  Here, exact results are not only available for thermal averages, but
for $\Omega(E)$ itself \cite{beale:96a}. Using the $K=0$ normalization condition
(\ref{omega_norm}), a single-histogram estimator for the density of states in the
energy language would be given by
\begin{equation}
  \hat{\Omega}(E) = 2^{\cal N}\,\frac{\hat{H}_K(E)\,e^{KE}}{\sum_E \hat{H}_K(E)\,e^{KE}}.
  \label{omega_est}
\end{equation}
This works quite well in the high-temperature phase and for small lattices. For lower
temperatures, however the histogram loses contact with the normalization point $K=0$,
resulting in large deviations from the correct normalization, cp.\ 
Fig.~\ref{omega_fig}. Clearly, each simulation samples only a rather small window of
energy space; from the exponential in the denominator of Eq.~(\ref{omega_est}),
however, configurations near the maximal energy $E = -{\cal N}$ receive the largest
weight in the sum, so that missing those configurations, which is the case for large
$K$, results in an exponentially wrong normalization factor (linear in $\ln
\Omega(E)$).  In other words, the absolute normalization condition (\ref{omega_norm})
reweights the histogram data to the point $K=0$, which will have no reliable outcome
if the overlap between the histograms at the simulation coupling and at $K=0$ is too
small or even vanishes. Note also, that the statistical error bars given in
Fig.~\ref{omega_fig} do not reflect this fundamental failure, although it is
statistical in nature. This is due to the fact that the usual implementation of error
estimation schemes for histograms take the error of histogram bins without entries to
be zero, whereas according to $1/N$ statistics it should in some sense be considered
infinitely large.

\begin{figure}[!tb]
  \unitlength8.75cm
  \begin{picture}(1,0.7)
    \put(-0.00886287625418057, -0.0313628762541806){\includegraphics[angle=0,clip=true,
      width=8.85535117056856cm,keepaspectratio=true]{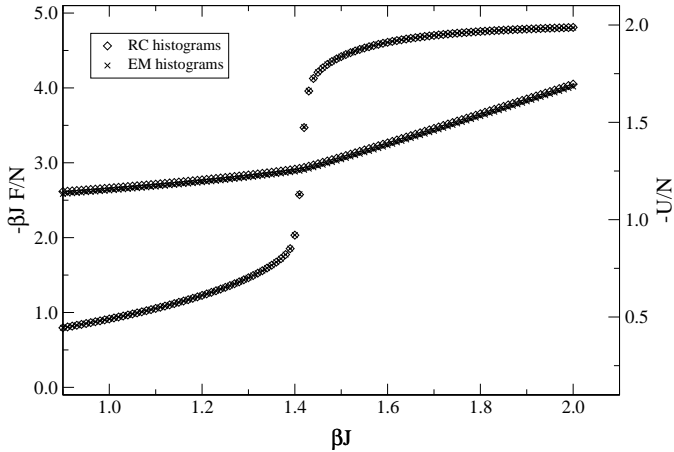}}
  \end{picture}
  \caption{Internal energy and free energy of the $q=10$ Potts model on a ${\cal
      N}=16^2$ square lattice as computed by the adaptively normalized RC
    multi-histograming scheme according to Eqs.~(\ref{selfcon_1}) and
    (\ref{selfcon_2}).  The results from the EM multi-histogram analysis of the same
    data are shown for comparison.}
  \label{q10_adapt_hu_sf}
\end{figure}

Expecting a similar sampling-related normalization failure for the RC histograms
normalized by the condition (\ref{sum_rule}) let us relax this absolute normalization
and apply an adaptive normalization scheme as in the original EM multi-histograming
formulation of Ref.~\cite{ferrenberg:89a}. Consider single-histogram estimates of the
partition function from simulations at $(p_i,q_i)$ and define reduced free energies
${\cal F}_i$ as
\begin{equation}
  F_i = -\frac{1}{K_i}\ln Z_{p_i,q_i}({\cal G}) = -{\cal E}-\frac{{\cal F}_i}{K_i}.
\end{equation}
From Eq.~(\ref{first_nt_est}) estimates of the density $g(b,n)$ from the single
histograms $\hat{H}_{p_i,q_i}(b,n)\equiv \hat{H}^{(i)}(b,n)$ are given by:
\begin{equation}
  \hat{g}^{(i)}(b,n) = e^{{\cal F}_i}\,\frac{\hat{H}^{(i)}(b,n)}{p_i^b\,(1-p_i)^{{\cal
        E}-b}\,q_i^n\,N_i}.
\end{equation}
Once again combining these estimates in a variance-optimized way as above in Section
II, treating the ${\cal F}_i$ as parameters with zero variance, we arrive at the
following expression
\begin{equation}
  \hat{g}(b,n) = \frac{\sum_i N_i\,e^{-{\cal F}_i}\,p_i^b\,(1-p_i)^{{\cal
        E}-b}\,q_i^n}
  {\sum_j N_j^2\,e^{-2{\cal F}_j}\,p_j^{2b}\,(1-p_j)^{2({\cal E}-b)}\,
    q_j^{2n}[\hat{H}^{(i)}(b,n)]^{-1}}.
  \label{selfcon_1}
\end{equation}
Now, from this estimate one has the following {\em a posteriori\/} relation for
computation of the parameters ${\cal F}_i$:
\begin{equation}
  e^{{\cal F}_i} = \sum_{b=0}^{\cal E}\sum_{n=1}^{\cal N}
  \hat{g}(b,n)\,p_i^b\,(1-p_i)^{{\cal E}-b}\,q_i^n.
  \label{selfcon_2}
\end{equation}
Eqs.~(\ref{selfcon_1}) and (\ref{selfcon_2}) form a pair of equations to be solved
self-consistently for the determination of the parameters ${\cal F}_i$, which can be
straightforwardly iterated by plugging in the results for ${\cal F}_i$ from
Eq.~(\ref{selfcon_1}) into Eq.~(\ref{selfcon_2}) and vice versa. One can improve on
that by applying more sophisticated iteration schemes like, e.g., the Newton-Raphson
iteration \cite{numrec}. We find, however, that the radius of convergence of this
method is quite small; therefore, we adaptively revert to the simple iteration if the
procedure leaves the Newton-Raphson convergence region. It is obvious that for the
iteration to converge, one needs some overlap between the $\hat{H}(b,n)$ histograms
between ``adjacent'' simulations, i.e., at least pairwise overlap. Apart from this
restriction, however, we find this iterative scheme to be very well-behaved,
converging rapidly in every case that fulfils the overlap-condition.

To get started, we use first-guess values of the ${\cal F}_i$ from thermodynamic
integration. Assume that the simulation points $(i)=(P_i,q_i)$ are ordered such
that the histograms of $(i)$ and $(i+1)$ have reasonable overlap; then
\begin{equation}
  e^{{\cal F}_i} = \sum_{b=0}^{\cal E}\sum_{n=1}^{\cal N}
  \frac{\hat{H}^{(i-1)}(b,n)}{N_{i-1}}\,\frac{p_i^b\,(1-p_i)^{{\cal E}-b}\,q_i^n}
  {p_{i-1}^b\,(1-p_{i-1})^{{\cal E}-b}\,q_{i-1}^n}\,e^{{\cal F}_{i-1}}
\end{equation}
is a good starting point for the described iteration scheme. ${\cal F}_1$ can be
chosen arbitrarily, since the given pair of equations is obviously invariant under a
global shift ${\cal F}_i \rightarrow {\cal F}_i - {\cal F}_1$. Thus, we have
determined the final estimate $\hat{g}(b,n)$ only up to a global factor. To fix this
last normalization we propose two different possibilities; on the one hand, we can
use the free model limit, i.e., evaluate ${\cal F}_{p=0,q}$ from
Eq.~(\ref{selfcon_2}) and use Eq.~(\ref{omega_norm}) for any $q$:
\begin{equation}
  {\cal F}_{p=0,q} = \ln Z_{p=0,q}({\cal G})-K{\cal E} = {\cal N}\ln q.
  \label{p0norm}
\end{equation}
On the other hand, also the $q=1$ partition function is trivial,
\begin{equation}
  {\cal F}_{p,q=1} = \ln\left[\sum_{b=0}^{\cal E} {{\cal E} \choose b}
    p^b\,(1-p)^{{\cal E}-b}\right] = 0,
  \label{q0norm}
\end{equation}
and can serve as normalization point for arbitrary $p$. In practice the best choice
depends on the set of simulated couplings $(p_i,q_i$): for large-$q$ simulations one
might want to resort to Eq.~(\ref{p0norm}), while otherwise Eq.~(\ref{q0norm}) should
be the better choice.

\begin{figure}[!t]
  \unitlength8.75cm
  \begin{picture}(1,0.7)
    \put(-0.0117056856187291, -0.0313628762541806){\includegraphics[angle=0,clip=true,
      width=8.19617474916388cm,keepaspectratio=true]{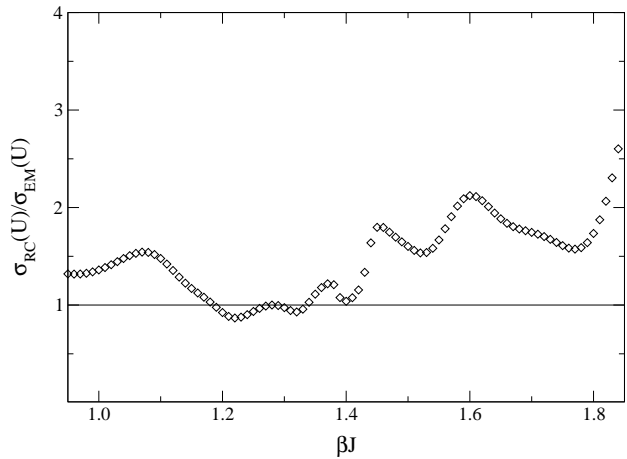}}
  \end{picture}
  \caption{Ratio of standard deviations for estimates of the internal energy $U$ 
    of the two-dimensional $q=10$ Potts model on a ${\cal N}=16^2$ lattice and RC and
    EM multi-histogram analyses as a function of the coupling $K=\beta J$. The
    variances are estimated by a ``jackknife'' time-series analysis \cite{efron:82}.
    The RC histograms are normalized according to Eqs.~(\ref{selfcon_1}) and
    (\ref{selfcon_2}).}
  \label{energy_error_q10}
\end{figure}

Now, we can re-consider the internal energy of the $q=10$ case from above with the
new, adaptively normalized RC multi-histograming scheme. Figure \ref{q10_adapt_hu_sf}
shows internal energy and free energy from this analysis as compared to the EM
multi-histogram approach. As far as the error estimates are concerned, we apply the
jackknife process to the whole iteration run, i.e., the iteration scheme for fixing
the weights ${\cal F}_i$ is done for each jackknife block of data separately, taking
full account of statistical errors. Clearly, now the results from both approaches
perfectly agree, the deviations of Fig.~\ref{q10_ising_hu_sf} have vanished. As
anticipated in Sec.~II, also the cluster estimator for the internal energy performs
noticeably better than in the $q=2$ case, such that --- at least in the critical region
--- it is quite comparable in precision to the EM estimator, cp.\ 
Fig.~\ref{energy_error_q10}.

For the free energy it is obvious that with the adaptive normalization scheme of RC
histograms we lose the especially high precision throughout the whole $K$ region
obtained by the application of the sum rule (\ref{sum_rule}) in
Fig.~\ref{ising_hu_sf} for the Ising model. To amend this, having fixed the relative
normalization of the single histograms adaptively, one might consider applying the
sum rule (\ref{sum_rule}) to the final result $\hat{g}(b,n)$ instead of using the
normalizations Eq.~(\ref{q0norm}) or Eq.~(\ref{p0norm}). This, however, gives results
looking almost identical to those shown above in Fig.~\ref{q10_hu_sf}, i.e., the
large deviations reappear, which clearly reveals the source they are resulting from.

\section{Comparison of the methods}

\begin{figure}[!t]
  \unitlength8.75cm
  \begin{picture}(1,0.7)
    \put(0.01, 0.0){\includegraphics[angle=0,clip=true,
      width=8.1cm,height=5.7cm]{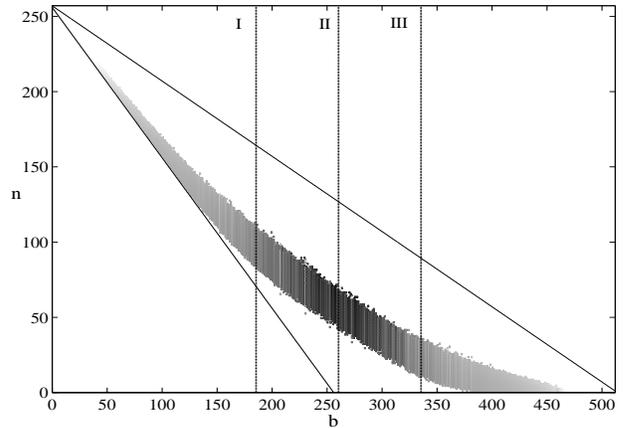}}
  \end{picture}
  \caption{Density plot of the relative differences
    $[\hat{g}^{(\mathrm{abs})}-\hat{g}^{(\mathrm{rel})}]/\hat{g}^{(\mathrm{rel})}$ of
    the density of states as sampled from the $q=10$ Potts model on a ${\cal N}=16^2$
    lattice by the absolutely normalized RC histograming scheme of Eq.~(\ref{nt_abs})
    ($\hat{g}^{(\mathrm{abs})}$) and the adaptively normalized scheme of
    Eq.~(\ref{selfcon_1}) ($\hat{g}^{(\mathrm{rel})}$), respectively. Dark shading
    indicates that $\hat{g}^{(\mathrm{abs})}(b,n) > \hat{g}^{(\mathrm{rel})}(b,n)$
    and vice versa.}
  \label{density_plot}
\end{figure}

The effect of this normalization problem should also be clearly seen in the final
estimates for the density of states $g(b,n)$ from the two RC histograming methods.
In Fig.~\ref{density_plot} we show a density plot of the relative differences of the
estimated density of states $\hat{g}(b,n)$ from the absolutely normalized
histograming scheme of Eq.~(\ref{nt_abs}), $\hat{g}^{(\mathrm{abs})}(b,n)$, and of
the adaptively normalized scheme of Eqs.~(\ref{selfcon_1}) and (\ref{selfcon_2}),
$\hat{g}^{(\mathrm{rel})}(b,n)$, i.e., the quantity
\begin{equation}
  \hat{\Delta}(b,n) \equiv \frac{\hat{g}^{(\mathrm{abs})}(b,n)-\hat{g}^{(\mathrm{rel})}(b,n)}
  {\hat{g}^{(\mathrm{rel})}(b,n)}.
\end{equation}
Note that the range of possible value pairs $(b,n)$ is restricted by two simple
bounds in the $(b,n)$ plane. First, starting from the point $(b=0,n={\cal N})$ each
added bond can at most reduce the number of clusters by one, namely by connecting two
previously unconnected clusters, i.e., one has
\begin{equation}
  n \ge {\cal N}-b.
\end{equation}
On the other hand, starting from the ``opposite'' point $(b={\cal E},n=1)$ one has
$b{\cal N}/{\cal E}$ bonds per site, so that for producing a new cluster one must at
least remove ${\cal N}/{\cal E}$ bonds,
\begin{equation}
  n-1 \le \frac{\cal N}{\cal E}({\cal E}-b),
\end{equation}
or, for the square lattice,
\begin{equation}
  n \le {\cal N}-\frac{b}{2}+1.
\end{equation}
Apart from single points near those bounds, all configurations within this triangle
can actually appear in a Potts model simulation with non-vanishing probability.

Now, from Fig.~\ref{density_plot} it is obvious, given that the estimate
$\hat{g}^{(\mathrm{rel})}(b,n)$ is correct up to an overall factor, that the
absolutely normalized histograming estimate $\hat{g}^{(\mathrm{abs})}(b,n)$ gives too
large estimates for $b$ values near the centre $b={\cal N}$ as compared to the other
regions of $b$ (dark shading in Fig.~\ref{density_plot}). Then, considering again the
deviation in internal energy shown in Fig.~\ref{q10_hu_sf}, its origin becomes clear:
the histogram $\hat{H}_{p,q}(b,n)$ for a simulation somewhat below the transition
point will be centred around the line $b=b_\mathrm{I}$ in Fig.~\ref{density_plot};
then, using the density of states estimate $\hat{g}^{(\mathrm{abs})}(b,n)$ for
evaluating $U$, the parts of the histogram lying to the right of $b=b_\mathrm{I}$
will have too large weight as compared to the values $b < b_\mathrm{I}$, thus by
Eq.~(\ref{hu_thermal}) resulting in a too large estimate for the internal energy $U$.
On the other hand, for couplings above the transition point the histogram will be
centred around $b=b_\mathrm{III}$ with too large weights for $b < b_\mathrm{III}$,
leading to estimates for $U$ that are too low. Directly in the vicinity of the
transition point, deviations in normalization are symmetric with respect to the
histogram, which will be centred around $b=b_\mathrm{II}$, thus leading to an
unbiased estimate for $U$. This is exactly the behavior found in
Fig.~\ref{q10_hu_sf}. Finally, contemplating on the reason for the deviations in
normalization shown in Fig.~\ref{density_plot} in the first place, it becomes obvious
that they have the same origin as those shown in Fig.~\ref{omega_fig}. The
exponential factor $q^{-n}$ from the sum rule Eq.~(\ref{sum_rule_data}) attaches
large weight to the configurations with small number of clusters $n$; if, however,
histograms miss entries for small $n$, as is the case for histograms in the
transition region $b \approx {\cal N}$ of Fig.~\ref{density_plot}, the sum
$\sum_n\hat{H}_{p,q}(b,n)\,q^{-n}$ will become too small, resulting in too large
normalization factors $\hat{C}_{p,q}(b)$.

\begin{figure}[!tb]
  \unitlength9cm
  \begin{picture}(1,0.7)
    \put(-0.0131270903010033, -0.0313628762541806){\includegraphics[angle=0,clip=true,
      width=8.44314381270903cm,keepaspectratio=true]{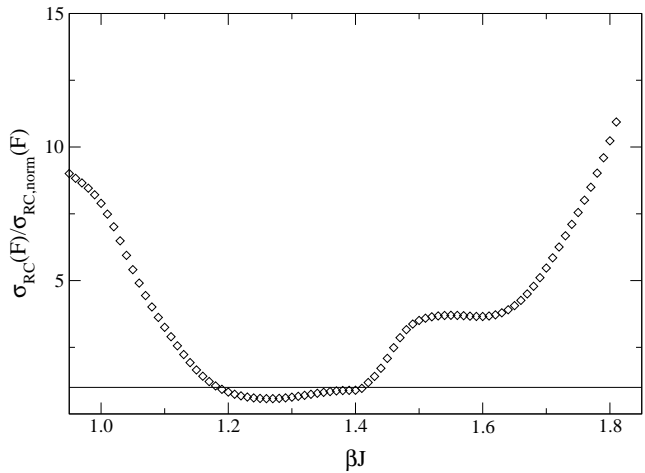}}
  \end{picture}
  \caption{Ratio of standard deviations of estimates of the free energy from
    the adaptively normalized RC histograming scheme of Eq.~(\ref{selfcon_1}) with
    ($\sigma_\mathrm{RC,norm}$) and without ($\sigma_\mathrm{RC}$) a final
    application of the sum rule Eq.~(\ref{sum_rule}) to the density of states after
    determining the weights. The used time series data includes both, the $q=2$ and
    the $q=10$ simulations reported above.}
  \label{q10_free_error}
\end{figure}

Thus, for the application of the sum rule (\ref{sum_rule}) to the final result from
the adaptively normalized RC histograming scheme to work reliably, one always has to
include histograms from small-$q$ simulations, like percolation ($q\rightarrow 1$) or
the Ising model ($q=2$), that produce configurations with relatively small numbers of
clusters $n$. To illustrate this, we combined the data from the $q=2$ and $q=10$
simulations reported above, used the histograming scheme Eqs.~(\ref{selfcon_1}) and
(\ref{selfcon_2}) to get results for $q=10$ and applied the sum rule
Eq.~(\ref{sum_rule}) afterwards, i.e., the normalization of the single histograms was
found adaptively, whereas the total histogram was normalized by the sum rule
(\ref{sum_rule}); this yields results for the internal and free energies
indistinguishable from those of the pure $q=10$ results of
Fig.~\ref{q10_adapt_hu_sf}. For the free energy, however, the size of statistical
errors is largely affected by the final normalization, cf.\ 
Fig.~\ref{q10_free_error}. For most of the couplings shown, the estimate from the
finally sum-rule-normalized density of states is up to about 10 times more accurate
in terms of the statistical errors. The presence and size of such a gain for a given
coupling is not mainly physically motivated, but rather depends on the relation of
the simulation points $(q_i,p_i)$ to the points of data analysis.
%
\enlargethispage*{1.5ex}
\section{Conclusions}

We have considered multi-histogram data analyses of time series from cluster-update
Monte Carlo simulations of the $q$-state Potts model in the random-cluster language.
Generalizing the original formulation of Ref.~\cite{hu:94a} to the case of
simulations of different number of states $q$, we found the original ansatz of
absolutely normalizing the individual histograms with a geometrical sum rule for
finite-length time series to produce large deviations from the expected behavior
when applied to cases $q$ larger then about $3$ or $4$ in two dimensions. We track
this error down to a mismatch between exponential suppression of a part of the
state space $(b,n)$ and a simultaneous exponential enhancement of this region in the
sum rule Eq.~(\ref{sum_rule_data}). To circumvent this problem, we propose a
different ansatz, normalizing the histograms adaptively via a set of self-consistency
equations aiming at the minimization of the variance of the final estimate of the
density of states $g(b,n)$. Absolute normalization over the whole temperature region
can still be maintained by making contact with the trivial partition function of the
percolation limit $q\rightarrow 1$ or by combining large- and small-$q$ data and
applying the sum rule (\ref{sum_rule}) after the adaptive normalization. This new
approach does not exhibit the limitations of the absolutely normalized ansatz to
small-$q$ simulations.

Comparing the newly introduced, adaptively normalized random-cluster ``RC''
multi-histogram technique with multi-histograming in the energy/magnetization ``EM''
language, we can make the following statements: (a) The cluster variables $(b,n)$
form the natural state space for the analysis of the Potts model. Using the
Swendsen-Wang cluster-update algorithm, these numbers are automatically known as a
by-product of the update-steps; no additional measurement steps are needed. (b) The
RC representation allows for reweighting in both parameters, the thermal coupling $p$
resp.\ $K$, and the number of states $q$, without systematical errors as in the
partial transformation of Ref.~\cite{lee:91a}.  Especially, the model can be
considered for the case of non-integer $q$. It is easy to combine data from
different-$q$ simulations to enhance the accuracy for large $q$.  (c) Cluster
estimators occur naturally in the RC language. Although we found that short-distance
observables like the internal energy and specific heat are sampled systematically
less acurate by cluster estimators, this situation is reversed for observables
sensitive to long-range order like the magnetization, susceptibility and correlation
functions.  Also, even short-range cluster estimators perform comparable to EM
language estimators for larger $q$ values, at least in the transition region. In the RC
language, the magnetic observables can be defined consistently throughout the broken
and unbroken phases, cf.\ Appendix A.

Apart from that, the combination of data from small and large $q$ models can serve as
a new method to cope with the super-critical slowing down at the first-order
transitions for large $q$: for sufficiently large lattices simulation runs will
entirely stay in one of the pure phases depending on the initial configuration and
boundary conditions due to ergodicity breaking at the transition point. However,
combining such data with smaller-$q$ simulations from the second-order or weak
first-order regime allows the adaptive normalization scheme to still find the correct
normalization of the pure-phase histograms without real tunneling events. This
approach is similar in spirit to the simulated tempering technique
\cite{marinari:92a,kerler:93a}.

One might also think of applying the multi-canonical \cite{berg:92b} resp.
multi-bondic \cite{wj:95a} simulation approach or one of the related techniques to
the sampling of the density of states $g(b,n)$. Especially, application of the
absolute normalization Eq.~(\ref{sum_rule}) to this case might be of interest. This
approach is currently under investigation. However, sampling the complete range of
possible values in the $(b,n)$ plane with sufficient accuracy is found to be a
computationally very demanding problem. In contrast, in the current approach, we
still stick to the physically sensible approach of importance-sampling, i.e.,
sampling the phase-space according to the local canonical weights. Furthermore, we
are able to take full advantage of the computational gain of cluster algorithms,
whereas the generalized-ensemble algorithms put forward so far employ local updates
(apart from the multi-bondic algorithm of Ref.~\cite{wj:95a}).

As an interesting application of our ansatz we suggest the analysis of the
tri-critical point $q_c$, where the order of the thermal transitions changes from
second to first order, in three dimensions. There has been quite some debate about
the location of this point, estimates ranging from $q_c = 2.15$ \cite{grollau:01a} to
$q_c = 2.6$ \cite{nienhuis:81a}. Furthermore, the universality class, critical
exponents etc.\ of this transition have not yet been properly analyzed.  A test for
the $q_c=4$ case in two dimensions shows that our method is well suited for such an
analysis.  This problem will be considered in a forthcoming publication.

\begin{acknowledgments}
  The collaboration of the authors has been supported by DAAD-NSC grant No.\
  D/9827248. MW acknowledges support by the DFG through the ``Graduiertenkolleg
  QFT''.
\end{acknowledgments}

\appendix

\section{Cluster estimators}

Consider the Potts model coupled to an external magnetic field $H$ with Hamiltonian
\footnote{Note that in Ref.~\cite{hu:84a} a different coupling between magnetic field
  and spins is employed.}
\begin{equation}
  {\cal H} = -J \sum_{\langle i,j\rangle}\delta(\sigma_i,\sigma_j) - H \sum_i \delta(\sigma_i,1).
  \label{potts_hamilton_field}  
\end{equation}
Then, the random-cluster representation of the partition function on a graph ${\cal
  G}$ consisting of ${\cal N}$ sites and ${\cal E}$ edges (bonds) is given by
\begin{eqnarray}
  Z_{p,q}({\cal G},B) & = & \sum_{{\cal G}' \subseteq {\cal G}} (e^K-1)^{b({\cal
      G}')}\prod_c\,[(q-1)+e^{Bn_c}] \nonumber \\
  & = & e^{K{\cal E}}\,\sum_{{\cal G}' \subseteq {\cal G}} 
  p^{b({\cal G}')}(1-p)^{{\cal E}-b({\cal G}')} \nonumber \\
  & & \times \prod_c\,[(q-1)+e^{Bn_c}],
\end{eqnarray}
where the product runs over the set of clusters $\{c\}$ of the subgraph ${\cal G}'$,
$n_c$ is the number of sites in cluster $c$, $K=\beta J$ is the thermal coupling
parameter and $B=\beta H$ denotes the reduced magnetic field. $p=1-e^{-K}$ is the
probability for the activation of bonds.

The zero-field, per-site internal energy $u$ is then given by
\begin{eqnarray}
  u & = & -\frac{\partial}{\partial K}\left[\frac{\ln Z_{p,q}({\cal G},B=0)}{{\cal
  N}}\right] \nonumber \\
  & = & -\frac{1}{\cal N}\left\langle\frac{\frac{\partial}{\partial K}
        [(e^K-1)^b]}{(e^K-1)^b}\right\rangle = -\frac{1}{p}\left\langle\frac{b}{\cal N}\right\rangle,
\end{eqnarray}
which shows the close connection between the $b$ and $E$ distributions.
\begin{widetext}
  The zero-field specific heat $c_v$ follows from
  \begin{eqnarray}
    c_v & = & K^2\frac{\partial^2}{\partial K^2}\left[\frac{\ln Z_{p,q}({\cal G},B=0)}{{\cal
          N}}\right] = \frac{K^2}{\cal N}\left[-\frac{1}{p^2}\langle
      b\rangle^2+\left\langle\frac{\frac{\partial^2}{\partial
          K^2}[(e^K-1)^b]}{(e^K-1)^b}\right\rangle\right] \nonumber \\
      & = & \frac{K^2}{p^2{\cal N}}\left[\langle b^2\rangle-\langle
          b\rangle^2-(1-p)\langle b\rangle\right].
  \end{eqnarray}
  
  In the thermodynamic limit, the per-site, zero-field ``magnetization''
  $\tilde{m}=\langle\sum_i\delta(\sigma_i,1)/{\cal N}\rangle$ is given by
  \begin{eqnarray}
    \tilde{m} & = & \lim_{B\rightarrow 0+}\,\lim_{{\cal N}\rightarrow\infty}\,
    \frac{\partial}{\partial B}\left[\frac{\ln Z_{p,q}({\cal G},B)}{{\cal N}}\right]
    \nonumber \\
    & = & \lim_{B\rightarrow 0+}\,\lim_{{\cal
        N}\rightarrow\infty}\,\frac{e^{K{\cal E}}}{Z_{p,q}({\cal G},B)}\,
    \sum_{{\cal G}' \subseteq {\cal G}}p^{b({\cal G}')}(1-p)^{{\cal E}-b({\cal G}')}
    \prod_c\,[(q-1)+e^{Bn_c}]\,\sum_{c'}\frac{n_c'}{\cal
      N}\,\frac{e^{Bn_c'}}{(q-1)+e^{Bn_c'}}
    \nonumber \\
    & = & \lim_{B\rightarrow 0+}\,\lim_{{\cal N}\rightarrow\infty}\,\left[
      \left\langle\sum_{c^{\pi}} \frac{n_c^{\pi}} {\cal
          N}\,\frac{1}{(q-1)e^{-Bn_c^{\pi}}+1}\right\rangle
      +\left\langle\sum_{c^{\phi}}\frac{n_c^{\phi}}{\cal
          N}\,\frac{1}{(q-1)e^{-Bn_c^{\phi}}+1}\right\rangle
    \right].
  \end{eqnarray}
\end{widetext}
Here, we split the cluster contributions of the subgraph into percolating clusters
$c^\pi$ and non-percolating, finite clusters $c^\phi$. In the indicated order of
taking the limits, first ${\cal N}\rightarrow\infty$ and then $B\rightarrow 0$, the
factors $\exp(-Bn_c)$ take the values $0$ and $1$ for percolating and non-percolating
clusters $c$, respectively. Thus, we arrive at
\begin{eqnarray}
  \tilde{m} & = & \left\langle\sum_{c^{\pi}}\frac{n_c^{\pi}}{\cal N}\right\rangle +
  \frac{1}{q}\,\left\langle\sum_{c^{\phi}}\frac{n_c^{\phi}}{\cal N}\right\rangle
  \nonumber \\ & = & \frac{q-1}{q}\left\langle\sum_{c^{\pi}}\frac{n_c^{\pi}}{\cal
      N}\right\rangle + \frac{1}{q},
\end{eqnarray}
which explicitly reflects the symmetry-breaking nature of the percolating
configurations. For the order parameter $m$, which varies between $0$ for the
completely disordered state and $1$ for the ground states, we find
\begin{equation}
  m \equiv \frac{q\tilde{m}-1}{q-1} = \left\langle\sum_{c^{\pi}}\frac{n_c^{\pi}}{\cal
      N}\right\rangle.
\end{equation}
For finite lattices one can retain this definition since the notion of percolating
and non-percolating clusters is still well-defined. Note, that this gives a
consistent definition of the order parameter throughout the disordered and broken
phases. In contrast, in the EM language one has to explicitly break symmetry in the
low-temperature phase, which is usually done by defining
\begin{equation}
  \tilde{m}_{K>K_t} = \left\langle\max_{1\le j\le q} \sum_i\delta(\sigma_i,j)\right\rangle,
  \label{m_broken}
\end{equation}
whereas for the unbroken phase one uses
\begin{equation}
  \tilde{m}_{K\le K_t} = \left\langle\sum_i\delta(\sigma_i,1)\right\rangle.
  \label{m_unbroken}
\end{equation}
Obviously, for finite lattices, the expectation values of the RC and EM definitions
will not coincide exactly; critical exponents, however, will of course agree.

The zero-field susceptibility $\tilde{\chi}$ is given by
\begin{widetext}
  \begin{eqnarray}
    \tilde{\chi} & = & \lim_{B\rightarrow 0+}\,\lim_{{\cal
        N}\rightarrow\infty}\,\frac{\partial^2}{\partial B^2}\left(\frac{\ln
        Z(B)}{{\cal N}}\right) = \lim_{B\rightarrow 0+}\,\lim_{{\cal
        N}\rightarrow\infty}\,\left[-\frac{1}{{\cal N} Z(B)^2}
      \left(\frac{\partial Z(B)}{\partial B}\right)^2 + 
      \frac{1}{{\cal N} Z(,B)}\frac{\partial^2 Z(B)}{\partial B^2}\right]
    \nonumber \\
        & = & -{\cal N} m^2 + \lim_{B\rightarrow 0+}\,\lim_{{\cal
        N}\rightarrow\infty}\left[\left\langle\sum_{c}\frac{n_c^2}{\cal
        N}\frac{1}{(q-1)e^{-Bn_c}+1}\right\rangle + 
    {\cal N}\left\langle\left(\sum_{c}\frac{n_c}{\cal
          N}\frac{1}{(q-1)e^{-Bn_c}+1}\right)^2\right\rangle\right.
    \nonumber \\
    & & \hspace{7.9cm} + \hspace{0.4cm}\left.\left\langle\sum_{c}\frac{n_c^2}{\cal
          N}\frac{1}{[(q-1)e^{-Bn_c}+1]^2}\right\rangle\right] \nonumber \\
    & = & {\cal
        N}\left(\frac{q-1}{q}\right)^2\left[\left\langle\left(\sum_{c^{\pi}}\frac{n_c^{\pi}}{\cal
        N}\right)^2\right\rangle-\left\langle\sum_{c^{\pi}}\frac{n_c^{\pi}}{\cal
        N}\right\rangle^2\right]+\frac{q-1}{q^2}\left\langle\sum_{c^{\phi}}\frac{{n_c^{\phi}}^2}{\cal
        N}\right\rangle.
  \label{cluster_sus}
  \end{eqnarray}
\end{widetext}
From Eq.~(\ref{cluster_sus}) one recognizes the widely used improved
cluster estimator for the high-temperature phase, namely the last term. Note,
however, that the original improved estimator includes {\em all\/} clusters here
instead of only the non-percolating ones, which makes a difference for finite
lattices. For finite lattices, once again, from Eq.~(\ref{cluster_sus}) we have a
single definition for both, the unbroken and broken phases.

Alternatively defining the susceptibility corresponding to the order parameter
$m$ we get
\begin{eqnarray}
  \chi & = & \left(\frac{q}{q-1}\right)^2\,\tilde{\chi} = \frac{1}{q-1}\left\langle\sum_{c^{\phi}}
    \frac{{n_c^{\phi}}^2}{\cal N}\right\rangle \nonumber \\
  & & + {\cal N}\left[\left\langle\left(\sum_{c^{\pi}}\frac{n_c^{\pi}}{\cal
          N}\right)^2\right\rangle-\left\langle\sum_{c^{\pi}}\frac{n_c^{\pi}}{\cal
        N}\right\rangle^2\right].
\end{eqnarray}


\begin{thebibliography}{27}
\expandafter\ifx\csname natexlab\endcsname\relax\def\natexlab#1{#1}\fi
\expandafter\ifx\csname bibnamefont\endcsname\relax
  \def\bibnamefont#1{#1}\fi
\expandafter\ifx\csname bibfnamefont\endcsname\relax
  \def\bibfnamefont#1{#1}\fi
\expandafter\ifx\csname citenamefont\endcsname\relax
  \def\citenamefont#1{#1}\fi
\expandafter\ifx\csname url\endcsname\relax
  \def\url#1{\texttt{#1}}\fi
\expandafter\ifx\csname urlprefix\endcsname\relax\def\urlprefix{URL }\fi
\providecommand{\bibinfo}[2]{#2}
\providecommand{\eprint}[2][]{\url{#2}}

\bibitem[{\citenamefont{Ferrenberg and Swendsen}(1988)}]{ferrenberg:88a}
\bibinfo{author}{\bibfnamefont{A.~M.} \bibnamefont{Ferrenberg}}
  \bibnamefont{and} \bibinfo{author}{\bibfnamefont{R.~H.}
  \bibnamefont{Swendsen}}, \bibinfo{journal}{Phys.\ Rev.\ Lett.}
  \textbf{\bibinfo{volume}{61}}, \bibinfo{pages}{2635} (\bibinfo{year}{1988});
  \bibinfo{journal}{Phys.\ Rev.\ Lett.}
  \textbf{\bibinfo{volume}{63}}, \bibinfo{pages}{1658(E)}
  (\bibinfo{year}{1989}{\natexlab{a}}).

\bibitem[{\citenamefont{Ferrenberg and
  Swendsen}(1989{\natexlab{b}})}]{ferrenberg:89a}
\bibinfo{author}{\bibfnamefont{A.~M.} \bibnamefont{Ferrenberg}}
  \bibnamefont{and} \bibinfo{author}{\bibfnamefont{R.~H.}
  \bibnamefont{Swendsen}}, \bibinfo{journal}{Phys.\ Rev.\ Lett.}
  \textbf{\bibinfo{volume}{63}}, \bibinfo{pages}{1195}
  (\bibinfo{year}{1989}{\natexlab{b}}).

\bibitem[{\citenamefont{Fortuin and Kasteleyn}(1972)}]{fortuin:72a}
\bibinfo{author}{\bibfnamefont{C.~M.} \bibnamefont{Fortuin}} \bibnamefont{and}
  \bibinfo{author}{\bibfnamefont{P.~W.} \bibnamefont{Kasteleyn}},
  \bibinfo{journal}{Physica} \textbf{\bibinfo{volume}{57}},
  \bibinfo{pages}{536} (\bibinfo{year}{1972}).

\bibitem[{\citenamefont{Coniglio and Klein}(1980)}]{coniglio:80a}
\bibinfo{author}{\bibfnamefont{A.}~\bibnamefont{Coniglio}} \bibnamefont{and}
  \bibinfo{author}{\bibfnamefont{W.}~\bibnamefont{Klein}},
  \bibinfo{journal}{J.\ Phys.\ A} \textbf{\bibinfo{volume}{13}},
  \bibinfo{pages}{2775} (\bibinfo{year}{1980}).

\bibitem[{\citenamefont{Hu}(1984)}]{hu:84a}
\bibinfo{author}{\bibfnamefont{C.-K.} \bibnamefont{Hu}},
  \bibinfo{journal}{Phys.\ Rev.\ B} \textbf{\bibinfo{volume}{29}},
  \bibinfo{pages}{5103} and \bibinfo{pages}{5109} (\bibinfo{year}{1984});
  \bibinfo{journal}{J.\ Phys.\ A} \textbf{\bibinfo{volume}{19}},
  \bibinfo{pages}{3067} (\bibinfo{year}{1986}).

\bibitem[{\citenamefont{Wolff}(1988)}]{wolff:88a}
\bibinfo{author}{\bibfnamefont{U.}~\bibnamefont{Wolff}},
  \bibinfo{journal}{Nucl.\ Phys.\ B} \textbf{\bibinfo{volume}{300}},
  \bibinfo{pages}{501} (\bibinfo{year}{1988}).

\bibitem[{\citenamefont{Hu}(1992{\natexlab{a}})}]{hu:92b}
\bibinfo{author}{\bibfnamefont{C.-K.} \bibnamefont{Hu}},
  \bibinfo{journal}{Phys.\ Rev.\ Lett.} \textbf{\bibinfo{volume}{69}},
  \bibinfo{pages}{2739} (\bibinfo{year}{1992}{\natexlab{a}}).

\bibitem[{\citenamefont{Chen and Hu}(1994)}]{hu:94a}
\bibinfo{author}{\bibfnamefont{J.-A.} \bibnamefont{Chen}} \bibnamefont{and}
  \bibinfo{author}{\bibfnamefont{C.-K.} \bibnamefont{Hu}},
  \bibinfo{journal}{Phys.\ Rev.\ B} \textbf{\bibinfo{volume}{50}},
  \bibinfo{pages}{6260} (\bibinfo{year}{1994}).

\bibitem[{\citenamefont{Swendsen and Wang}(1987)}]{swendsen-wang:87a}
\bibinfo{author}{\bibfnamefont{R.~H.} \bibnamefont{Swendsen}} \bibnamefont{and}
  \bibinfo{author}{\bibfnamefont{J.-S.} \bibnamefont{Wang}},
  \bibinfo{journal}{Phys.\ Rev.\ Lett.} \textbf{\bibinfo{volume}{58}},
  \bibinfo{pages}{86} (\bibinfo{year}{1987}).

\bibitem[{\citenamefont{Hu}(1992{\natexlab{b}})}]{hu:92a}
\bibinfo{author}{\bibfnamefont{C.-K.} \bibnamefont{Hu}},
  \bibinfo{journal}{Phys.\ Rev.\ B} \textbf{\bibinfo{volume}{46}},
  \bibinfo{pages}{6592} (\bibinfo{year}{1992}{\natexlab{b}}).

\bibitem[{\citenamefont{Ferdinand and Fisher}(1969)}]{ferdinand:69a}
\bibinfo{author}{\bibfnamefont{A.~E.} \bibnamefont{Ferdinand}}
  \bibnamefont{and} \bibinfo{author}{\bibfnamefont{M.~E.}
  \bibnamefont{Fisher}}, \bibinfo{journal}{Phys.\ Rev.}
  \textbf{\bibinfo{volume}{185}}, \bibinfo{pages}{832} (\bibinfo{year}{1969}).

\bibitem[{\citenamefont{Efron}(1982)}]{efron:82}
\bibinfo{author}{\bibfnamefont{B.}~\bibnamefont{Efron}},
  \emph{\bibinfo{title}{The Jackknife, the Bootstrap and Other Resampling
  Plans}} (\bibinfo{publisher}{Society for Industrial and Applied Mathematics
  [SIAM]}, \bibinfo{address}{Philadelphia}, \bibinfo{year}{1982}).

\bibitem[{\citenamefont{Onsager}(1944)}]{onsager:44}
\bibinfo{author}{\bibfnamefont{L.}~\bibnamefont{Onsager}},
  \bibinfo{journal}{Phys.\ Rev.} \textbf{\bibinfo{volume}{65}},
  \bibinfo{pages}{117} (\bibinfo{year}{1944}).

\bibitem[{\citenamefont{Wu}(1982)}]{wu:82a}
\bibinfo{author}{\bibfnamefont{F.~Y.} \bibnamefont{Wu}},
  \bibinfo{journal}{Rev.\ Mod.\ Phys.} \textbf{\bibinfo{volume}{54}},
  \bibinfo{pages}{235} (\bibinfo{year}{1982}).

\bibitem[{\citenamefont{Kaufman}(1949)}]{kaufman:49a}
\bibinfo{author}{\bibfnamefont{B.}~\bibnamefont{Kaufman}},
  \bibinfo{journal}{Phys.\ Rev.} \textbf{\bibinfo{volume}{76}},
  \bibinfo{pages}{1232} (\bibinfo{year}{1949}).

\bibitem[{\citenamefont{Wolff}(1990)}]{wolff:90a}
\bibinfo{author}{\bibfnamefont{U.}~\bibnamefont{Wolff}},
  \bibinfo{journal}{Nucl.\ Phys.\ B} \textbf{\bibinfo{volume}{334}},
  \bibinfo{pages}{581} (\bibinfo{year}{1990}).

\bibitem[{\citenamefont{Janke and Kappler}(1995)}]{wj:95a}
\bibinfo{author}{\bibfnamefont{W.}~\bibnamefont{Janke}} \bibnamefont{and}
  \bibinfo{author}{\bibfnamefont{S.}~\bibnamefont{Kappler}},
  \bibinfo{journal}{Phys.\ Rev.\ Lett.} \textbf{\bibinfo{volume}{74}},
  \bibinfo{pages}{212} (\bibinfo{year}{1995}).

\bibitem[{\citenamefont{Berg and Neuhaus}(1992)}]{berg:92b}
\bibinfo{author}{\bibfnamefont{B.~A.} \bibnamefont{Berg}} \bibnamefont{and}
  \bibinfo{author}{\bibfnamefont{T.}~\bibnamefont{Neuhaus}},
  \bibinfo{journal}{Phys.\ Rev.\ Lett.} \textbf{\bibinfo{volume}{68}},
  \bibinfo{pages}{9} (\bibinfo{year}{1992}).

\bibitem[{\citenamefont{Janke and Kappler}(1997)}]{wj:97a}
\bibinfo{author}{\bibfnamefont{W.}~\bibnamefont{Janke}} \bibnamefont{and}
  \bibinfo{author}{\bibfnamefont{S.}~\bibnamefont{Kappler}},
  \bibinfo{journal}{J.\ Phys.\ I (France)} \textbf{\bibinfo{volume}{7}},
  \bibinfo{pages}{663} (\bibinfo{year}{1997}).

\bibitem[{\citenamefont{Beale}(1996)}]{beale:96a}
\bibinfo{author}{\bibfnamefont{P.~D.} \bibnamefont{Beale}},
  \bibinfo{journal}{Phys.\ Rev.\ Lett.} \textbf{\bibinfo{volume}{76}},
  \bibinfo{pages}{78} (\bibinfo{year}{1996}).

\bibitem[{\citenamefont{Press et~al.}(1992)\citenamefont{Press, Teukolsky,
  Vetterling, and Flannery}}]{numrec}
\bibinfo{author}{\bibfnamefont{W.~H.} \bibnamefont{Press}},
  \bibinfo{author}{\bibfnamefont{S.~A.} \bibnamefont{Teukolsky}},
  \bibinfo{author}{\bibfnamefont{W.~T.} \bibnamefont{Vetterling}},
  \bibnamefont{and} \bibinfo{author}{\bibfnamefont{B.~P.}
  \bibnamefont{Flannery}}, \emph{\bibinfo{title}{Numerical Recipes in C --- The
  Art of Scientific Computing}} (\bibinfo{publisher}{Cambridge University
  Press}, \bibinfo{address}{Cambridge}, \bibinfo{year}{1992}),
  \bibinfo{edition}{2nd} ed.

\bibitem[{\citenamefont{Lee and Kosterlitz}(1991)}]{lee:91a}
\bibinfo{author}{\bibfnamefont{J.}~\bibnamefont{Lee}} \bibnamefont{and}
  \bibinfo{author}{\bibfnamefont{J.~M.} \bibnamefont{Kosterlitz}},
  \bibinfo{journal}{Phys.\ Rev.\ B} \textbf{\bibinfo{volume}{43}},
  \bibinfo{pages}{1268} (\bibinfo{year}{1991}).

\bibitem[{\citenamefont{Marinari and Parisi}(1992)}]{marinari:92a}
\bibinfo{author}{\bibfnamefont{E.}~\bibnamefont{Marinari}} \bibnamefont{and}
  \bibinfo{author}{\bibfnamefont{G.}~\bibnamefont{Parisi}},
  \bibinfo{journal}{Europhys.\ Lett.} \textbf{\bibinfo{volume}{19}},
  \bibinfo{pages}{451} (\bibinfo{year}{1992}).

\bibitem[{\citenamefont{Kerler and Weber}(1993)}]{kerler:93a}
\bibinfo{author}{\bibfnamefont{W.}~\bibnamefont{Kerler}} \bibnamefont{and}
  \bibinfo{author}{\bibfnamefont{A.}~\bibnamefont{Weber}},
  \bibinfo{journal}{Phys.\ Rev.\ B} \textbf{\bibinfo{volume}{47}},
  \bibinfo{pages}{11\,563} (\bibinfo{year}{1993}).

\bibitem[{\citenamefont{Grollau et~al.}(2000)\citenamefont{Grollau, Rosinberg,
  and Tarjus}}]{grollau:01a}
\bibinfo{author}{\bibfnamefont{S.}~\bibnamefont{Grollau}},
  \bibinfo{author}{\bibfnamefont{M.~L.} \bibnamefont{Rosinberg}},
  \bibnamefont{and} \bibinfo{author}{\bibfnamefont{G.}~\bibnamefont{Tarjus}},
  \bibinfo{howpublished}{cond-mat/0011483 preprint} (\bibinfo{year}{2000}).

\bibitem[{\citenamefont{Nienhuis et~al.}(1981)\citenamefont{Nienhuis, Riedel,
  and Schick}}]{nienhuis:81a}
\bibinfo{author}{\bibfnamefont{B.}~\bibnamefont{Nienhuis}},
  \bibinfo{author}{\bibfnamefont{E.~K.} \bibnamefont{Riedel}},
  \bibnamefont{and} \bibinfo{author}{\bibfnamefont{M.}~\bibnamefont{Schick}},
  \bibinfo{journal}{Phys.\ Rev.\ B} \textbf{\bibinfo{volume}{23}},
  \bibinfo{pages}{6055} (\bibinfo{year}{1981}).

\end{thebibliography}

\end{document}